\documentclass[aps,pra,11pt]{revtex4-1}
\usepackage{graphics}
\usepackage{color}
\begin{document}
\title{Predictability of Extreme Intensity Pulses in Optically Injected Semiconductor Lasers}
\author{Nuria Martinez Alvarez}
\affiliation{Departament de F\'{i}sica, Universitat Politecnica de Catalunya, Colom 11, 08222 Terrassa, Barcelona, Spain}

\author{Saurabh Borkar}
%\affiliation{Indian Institute of Technology, Guwahati, Assam, India}
\affiliation{Departament de F\'{i}sica, Universitat Politecnica de Catalunya, Colom 11, 08222 Terrassa, Barcelona, Spain}

\author{Cristina Masoller}
\affiliation{Departament de F\'{i}sica, Universitat Politecnica de Catalunya, Colom 11, 08222 Terrassa, Barcelona, Spain}
\email{cristina.masoller@upc.edu} 

%
%\section{Abstract}
\begin{abstract}
The predictability of extreme intensity pulses emitted by an optically injected semiconductor laser is studied numerically, by using a well-known rate equation model. We show that symbolic ordinal time-series analysis allows to identify the patterns of intensity oscillations that are likely to occur before an extreme pulse. The method also gives information about patterns which are unlikely to occur before an extreme pulse. The specific patterns identified capture the topology of the underlying chaotic attractor and depend on the model parameters. The methodology proposed here can be useful for analyzing data recorded from other complex systems that generate extreme fluctuations in their output signals. 
\end{abstract}
\date{\today}
\maketitle

\section{Introduction}
\label{intro}
Extreme events are ubiquitous in complex systems \cite{book,solli_2007,review,ulrike,klaus,theo} and a lot of efforts are nowadays focused on developing reliable analysis techniques for their detection and prediction \cite{book0,kantz,theo2,alam_2014,prl_2015,deluca,gauthier}. In many scientific fields, if an extreme event occurs in an unexpected way, it often has disastrous consequences. Examples include tsunamis, market crashes, earthquakes, population extinctions, etc. \cite{sornette,book2,davidsen}. For building safer environments and for developing appropriate mitigation strategies \cite{ipcc}, it is crucial to identify early-warnings of such events. Laser systems displaying extreme pulses in their output intensity \cite{pisarchik,meucci,Bonatto_PRL_2011,Zamora_PRA_2013,pra_2014,tredicce,sergei_nat_phot,prl_2016} are ideal candidates for performing laboratory controlled experiments that allow testing novel diagnostic tools to detect warning signals of upcoming extreme events. 

Here we propose a novel technique, based on symbolic ordinal time series analysis \cite{bandt2002permutation,rosso,zanin}.  We show, through simulations of a laser system known to display extreme pulses, that ordinal analysis detects the patterns of oscillations of the intensity pulses that more frequently (or less frequently) occur before an extreme pulse. The definition of ``extreme'' intensity pulses (which have been referred to as optical rogue waves) depends on the particular system under investigation. In hydrodynamics, when the height of a wave is larger by a factor of 3 than average, this wave is considered extreme. For example, regarding the wave height of tsunamis, maxima several times higher than the average wave height have been reported \cite{theo}. In optics, however, fluctuations of much higher amplitudes compared to the average are often observed \cite{prl_2016}. Here, we consider an optically injected semiconductor laser as in \cite{Bonatto_PRL_2011,Zamora_PRA_2013,pra_2014} and the thresholds used are in units of the intensity standard deviation.

\section{Model and method of analysis}

The rate equations describing the dynamics of a continuous-wave (cw) optically injected semiconductor laser are \cite{sebastian,gallas}
\begin{eqnarray}
  \frac{dE}{dt} &=& \kappa(1+i\alpha)(N-1)E +i\Delta\omega E + \sqrt{P_{inj}}+ \sqrt{D}\xi(t)  \\
  \frac{dN}{dt} &=& \gamma_N(\mu(t)-N-|E|^2)
\end{eqnarray}
where $E$ is the slow envelope of the complex electric field, $N$ is the carrier density, $\kappa$ is the field decay rate, $\alpha$ is the line-width enhancement factor, and $\gamma_N$ is the carrier decay rate. $\Delta \omega$ is the frequency detuning between the master and the injected laser and $P_{inj}$ is the injection strength. $\xi(t)$ is a complex Gaussian white noise of strength $D$ representing spontaneous emission. $\mu(t)=\mu_0+\mu_{mod}\sin(\omega_{mod} t)$ is the time-dependent injection current parameter (normalized such that the threshold of the free-running laser is at $\mu_{th}$=1), which is sinusoidally modulated: $\mu_0$ is the dc bias current, $\mu_{mod}$ is the modulation amplitude, and $\omega_{mod}$ is the modulation frequency.

The oscillations in the intensity time series are analyzed by using the symbolic method known as \textit{ordinal analysis} \cite{bandt2002permutation}. Within this framework, a time series $y(t)$ is divided into non-overlapping segments of length $L$, and each segment is assigned a symbol, $s$, (known as ordinal pattern, OP) according to the ranking of the values inside the segment. For example, with $L=3$, if $y(t) < y(t+1) < y(t+2)$, $s(t)$ is `012', if $y(t) > y(t+1) > y(t+2)$, $s(t)$ is `210', and so forth. Thus, the symbols take into account the \textit{relative temporal ordering} of the values and not the values themselves. In this way, each symbol encodes information about the ordering of $L$ consecutive data points. 

We apply the ordinal method to the sequence of intensity peak heights (see Fig. 1), $\{\dots I_{i}, I_{i+1}, I_{i+2} \dots \}$. For each peak that is above a given threshold, $I_0$, the $L=3$ previous consecutive peaks are used to define the ordinal patterns. In other words, if $I_{i}>I_0$, we study the three previous peaks, $\{I_{i-3}, I_{i-2}, I_{i-1}\}$: $I_{i-3} < I_{i-2}< I_{i-1}$ gives `$012$', $I_{i-2} < I_{i-3}< I_{i-1}$ gives `$120$', $I_{i-1} < I_{i-2}< I_{i-3}$ gives `$210$', etc. 

In order to estimate the OP probabilities with good statistics, long intensity time series were simulated, which had, for the maximum threshold considered, more than 1000 intensity pulses higher than the threshold. The interval of probability values consistent with the uniform distribution (the null hypothesis is that no preferred or infrequent patterns occur before a pulse higher than a given threshold $I_0$) was computed as $p \pm 3 \sigma_{p}$ (99.74\% confidence level), with $p=1/N_{OP}$ and $\sigma_{p} = \sqrt{p(1-p)/N}$, where $N_{OP}=6$ is the number of possible $L=3$ ordinal patterns and $N$ is the number of peak intensities $I_{i}>I_0$.

\section{Results}

The model equations were numerically solved using the same parameters as in \cite{Bonatto_PRL_2011,Zamora_PRA_2013,pra_2014}: $\kappa=300$ n$s^{-1}$, $\alpha=3$, $\gamma_N=1$ ns$^{-1}$,  $P_{inj}=60$ ns$^{-2}$, $\mu_0=2.4$, and the other parameters are indicated in the figure captions. Time traces of 20 $\mu$s were generated from random initial conditions. The bifurcation diagram of the deterministic model without pump current modulation is presented in Fig. 2. We consider as control parameter the frequency detuning. To do the bifurcation diagram, for each detuning the simulated intensity time series, $I'(t)=|E(t)|^2$, was normalized to remove the mean and to unit variance (i.e., $I(t)=[I'(t)-\left<I'\right>]/\sigma_{I'}$ where $\left<I'\right>$ and $\sigma^2_{I'}$ are the mean and the variance of $I'(t)$, respectively). Then, the sequence of $N$ consecutive peak intensities, $\{I_1 \dots I_{i} \dots I_N\}$ [indicated with dots in Figs. 1(a) and 1(b)] was plotted vs. the detuning. In this diagram high pulses occur in a narrow parameter region, in good agreement with \cite{Bonatto_PRL_2011,Zamora_PRA_2013,pra_2014}. In the following we analyze the intensity dynamics in this region. We also consider a second set of parameters, such that high intensity pulses are generated by the interplay of noise and current modulation \cite{pra_2014}.  

Figure 1 presents examples of the intensity time series. We observe that the dynamics is highly irregular and occasionally, very high pulses occur. Figure 3 displays the variation of the number of peak intensities that are higher than a given threshold, $I_0$, with the threshold $I_0$. Here we observe a monotonic threshold dependency: for low thresholds there are a large number of pulses $I_i>I_0$, but as the threshold increases, the number of pulses $I_i>I_0$ decreases fast (an almost linear variation in log vertical scale).  

For each peak above a given threshold, the three previous consecutive peaks were considered for defining the ordinal patterns, and their probability of occurrence was calculated for different thresholds. The results are displayed in Fig. 4, where the thick lines indicate the 99.74\% confidence interval that the OP probabilities are consistent with the uniform distribution. We note that the width of the confidence interval increases with the threshold because the number, $N$, of peak intensities that are above the threshold decreases as the threshold increases.

In Fig. 4(a), corresponding to deterministic simulations with no current modulation, it can be seen that pattern `201' occurs before each intensity peak whose height is above 5. In Fig. 4(b), corresponding to stochastic simulations with current modulation, no statistically significant frequent pattern occurs before the high pulses; however, before pulses higher than 6, pattern `210' occurs with about 60\% probability, while patterns `012', `021' and `102' never occur. An inspection of the time series displayed in Figs. 1(a), 1(b) confirms these observations. 

Interestingly, the more frequently observed ordinal pattern in the absence of noise and modulation (`201', squares) becomes an infrequent pattern in the presence of noise and modulation, at least for high values of the threshold. In addition, the pattern `210' (triangles), which, in the deterministic simulations, does not occur for high thresholds, becomes, in the simulations with noise and modulation, increasingly frequent with increasing threshold. The reasons for these variations are unclear and might be due to a change in the topology of the attractor.

\begin{figure}
\resizebox{1.0\columnwidth}{!}{\includegraphics{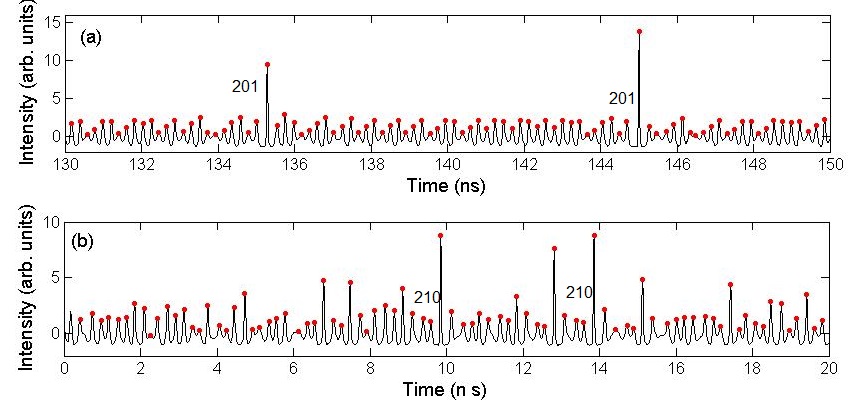} }
\caption{Intensity time-series normalized to zero-mean and unit variance. (a) Deterministic simulations with no modulation, the parameters are: $D=0$, $\mu_{mod}=0$, $\Delta\nu=\Delta \omega/2\pi=0.22$ GHz; (b) Stochastic simulations with current modulation, the parameters are: $D=0.001$ ns$^{-1}$, $\Delta\nu=\Delta \omega/2\pi=-0.24$ GHz, $\mu_{mod}=0.2$ and $f_{mod}=\omega_{mod}/2\pi=3$ GHz. In each panel two ordinal patterns occurring before high pulses are indicated.}
\label{fig:2}       
\end{figure}

\begin{figure}
\centering\resizebox{0.75\columnwidth}{!}{\includegraphics{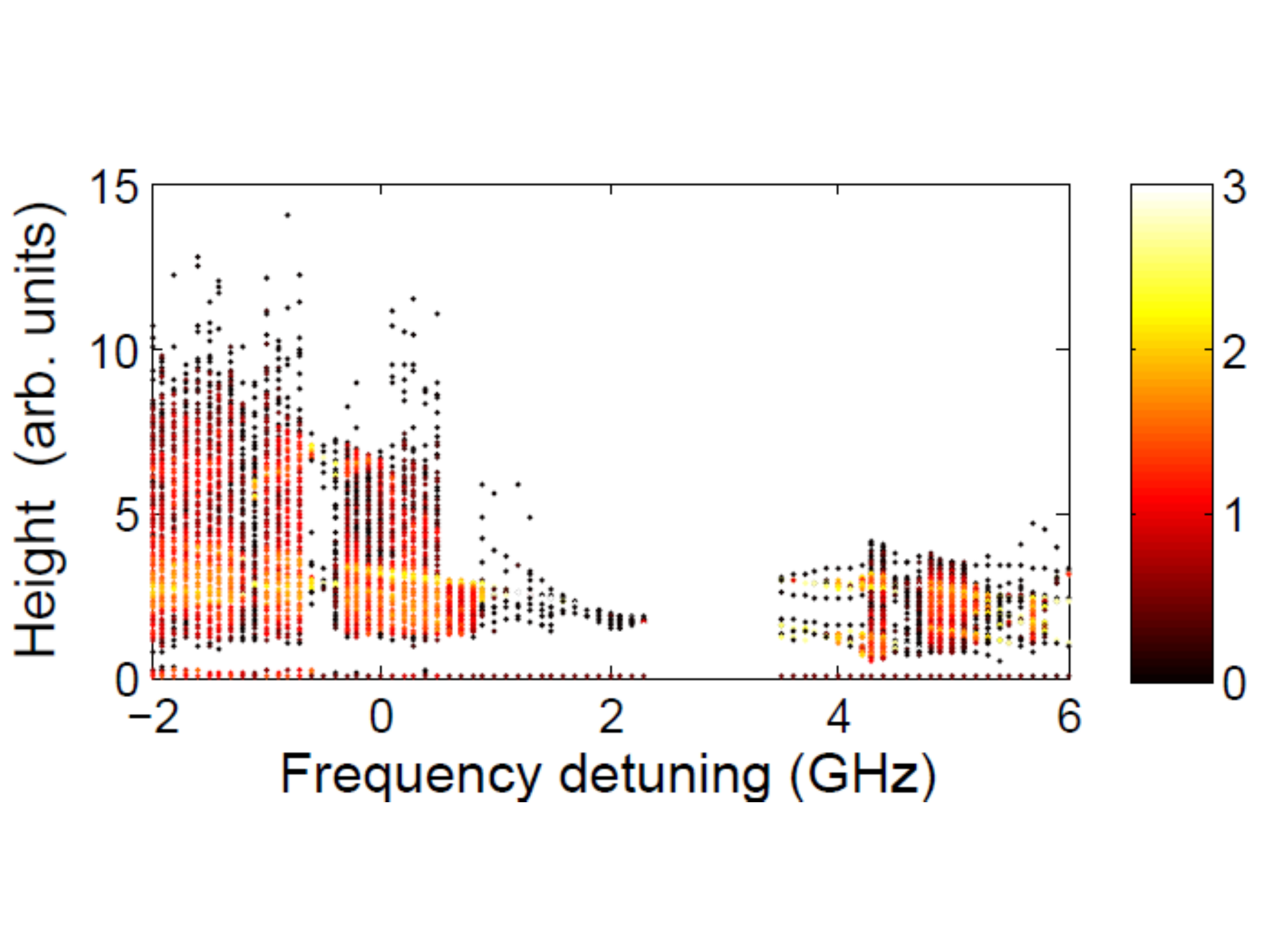} }
\caption{Bifurcation diagram displaying the height of the intensity oscillations vs the frequency detuning. The color code indicates $\log(N)$, where $N$ is the number of oscillations at a given height. The parameters are as indicated in the text.}
\label{fig:1}       
\end{figure}

\begin{figure}
\resizebox{1.0\columnwidth}{!}{\includegraphics{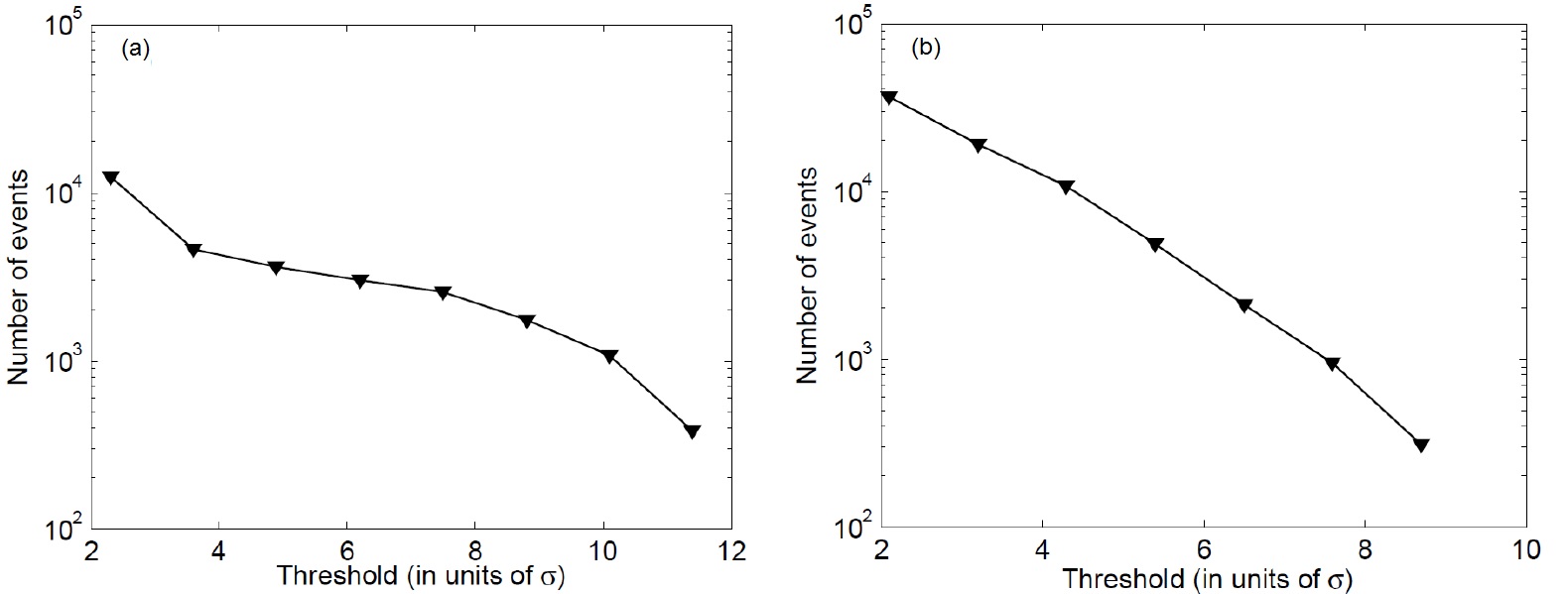} }
\caption{Number of peak intensities higher than the threshold (i.e., number of threshold-crossing events) vs. the threshold.  (a) Deterministic simulations with no modulation, (b) stochastic simulations with current modulation. The parameters are as in Fig. 1.}
\label{fig:3}       
\end{figure}

\begin{figure}
\resizebox{1.0\columnwidth}{!}{\includegraphics{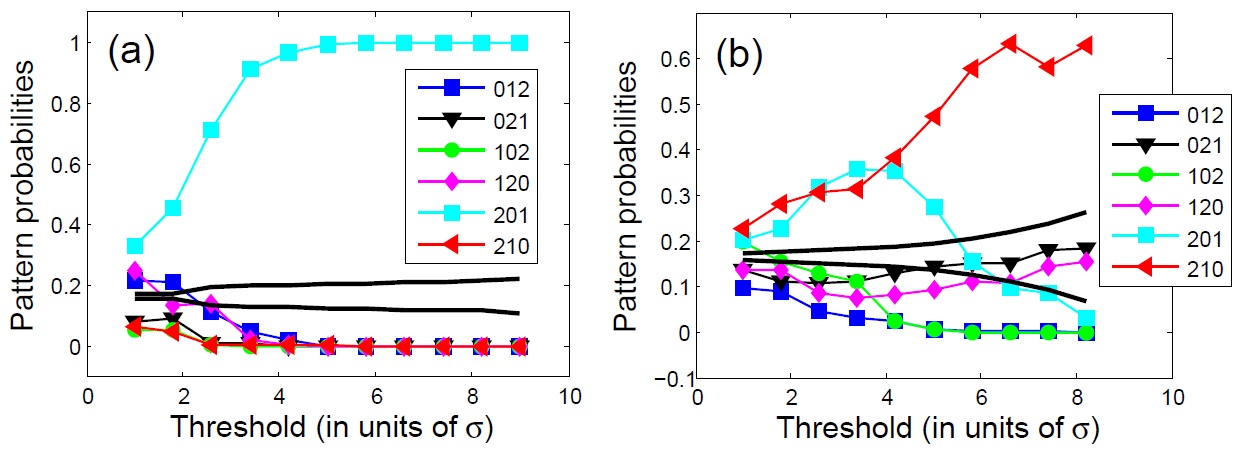} }
\caption{Probabilities of the ordinal patterns (OPs) formed by the three intensity peaks that occur before each intensity peak that is above a given threshold $I_0$ [i.e., for each $I_i>I_0$, the OP is defined by $\{I_{i-3}, I_{i-2}, I_{i-1}\}$] as a function of the threshold $I_0$. The thick lines indicate the range of probability values consistent with the uniform distribution, which is estimated with a binomial test. (a) Deterministic simulations with no modulation; (b) stochastic simulations with current modulation. The parameters are as in Fig. 1.}
\label{fig:4}      
\end{figure}

\section{Discussion and conclusions}

Extensive simulations show that our findings are robust, in the sense that, in the parameter regions where the intensity dynamics displays extreme pulses, these tend to be anticipated by well defined patterns of intensity oscillations; however, the specific pattern varies with the model parameters. This is due to the fact that the ordinal probabilities capture the topology of the underlying attractor, in the region of the attractor that is rarely visited, where extreme pulses are generated. We have also found model parameters for which no preferred $L=3$ ordinal pattern could be detected to occur before the extreme pulses; however, in these cases, by considering longer patterns defined by four or more intensity oscillations, we can in general identify more probable and less probable patterns that occur before the extreme pulses. A drawback of considering longer ordinal patterns is that they increase the data requirements, i.e., in order to compute their probabilities with robust statistics, a much larger number of extreme pulses is needed.

Another issue is the role of noise, dynamical and observational. If the intensity dynamics is too noisy (as in experimental data), in order to identify clear patterns of oscillations that anticipate the extreme pulses, the data should be preprocessed by applying an appropriate filter. A detailed study of the optimal pattern length and of the role of noise will be reported elsewhere. 

The methodology proposed here is expected to work whenever the underlying mechanisms generating the extreme fluctuations have deterministic components. We don't expect to gain useful insight with this approach if the extreme pulses are the result of random dynamics. 

The laser system considered here is an ideal candidate for a proof of principle of the predictive power of the ordinal approach because previous work has shown, experimentally and numerically, that the knowledge of the intensity as a function of time is enough to predict the appearance of an extreme pulse some time before it actually happens \cite{Zamora_PRA_2013}. It was also shown that the higher the threshold is, the longer the time in advance that the extreme pulse can be predicted.

The predicting power of the ordinal approach is limited in the sense that only the occurrence of a pulse above a given threshold can be forecasted with a certain probability, but not the actual height of the upcoming pulse. Other analysis techniques, such as the Grassberger-Procaccia algorithm \cite{gp} as used in \cite{prl_2015}, provide more information and can be used to improve predictability.

Concluding, we have shown that symbolic ordinal analysis can be a valuable diagnostic tool for identifying warnings of extreme intensity pulses in the form of the oscillation patterns that are most likely to occur before an extreme pulse. The method also gives information about patterns which are unlikely to occur before an extreme pulse, thus giving insight into ``safe'' time intervals. This approach could be valuable for studying the predictability of extreme optical pulses emitted by other laser systems \cite{tredicce,sergei_nat_phot,prl_2016}. Future work is aimed at testing this technique with optical experimental data and also, with data recorded from other systems that generate extreme fluctuations in their output signals.

\section{Acknowledgments}

This work was supported in part by Spanish MINECO (FIS2015-66503-C3-2-P) and ICREA ACADEMIA, Generalitat de Catalunya.

\end{document}